# Effect of a plasma grating on pump-probe experiments near the ionization threshold in gases

J. K. Wahlstrand and H. M. Milchberg

*Institute for Research in Electronics and Applied Physics, University of Maryland, College Park, MD 20742 USA*

Calculations are performed of the phase shift caused by the spatial modulation in the plasma density due to interference between a strong pump pulse and a weak probe pulse. It is suggested that a recent experiment [Loriot *et al.*, Opt. Express **17**, 13429 (2009)] observed an effective birefringence from this plasma grating rather than from the higher-order Kerr effect.

Recently, a saturating, then negative induced birefringence was measured at high intensities in Ar, $N_2$, and $O_2$ gas [1-3]. It was interpreted as the saturation followed by sign change of the instantaneous nonlinear response, and was described using higher order terms in a series expansion of the nonlinear refractive index. This has led to a slew of theoretical papers discussing the consequences of the higher-order Kerr effect (HOKE), specifically a strong negative nonlinear refractive index below the ionization threshold [4-7]. Many experiments have since investigated HOKE, mostly concentrating on its effects on phenomena such as filamentation, harmonic generation, and conical emission [8-14]. A recent measurement of the nonlinearity using spectral interferometry showed no sign of HOKE up to and beyond the ionization threshold [15], in disagreement with the results of [1-3]. Here, a possible source of the discrepancy between these two experiments is described. We find that when the pump intensity is near the ionization threshold, a refractive index grating formed by the interference of degenerate pump and probe beams produces a negative phase shift in the probe pulse, and can effectively act as a birefringent medium.

In the pump-probe transient birefringence experiment considered [1-3], a strong pump beam and a weak probe beam, both linearly polarized, are focused in a gas cell, crossing at a 4° angle. The relative angle between the polarization of the pump and probe is 45°, and the transient birefringence is measured using an analyzer oriented to block the probe in the absence of a birefringence. The heterodyne signal, measured by adding and subtracting a small birefringence using a static phase plate [1], is linear in the birefringence induced in the medium by the pump. The optical Kerr effect produces a birefringence because, in an isotropic medium, the Kerr coefficient is 3 times as large for the probe polarization parallel to the pump polarization as it is for the perpendicular component. It had been assumed that plasma, because it produces a change in refractive index that is isotropic, cannot produce a birefringence [1,2].

Here we show that the plasma grating created due to interference between pump and probe beams can produce an effective birefringence that appears only when the pump and probe pulses overlap in time.

For incident pump and probe fields $\mathbf{E}_e(\mathbf{r},t) = \hat{\mathbf{x}} E_e(\mathbf{r},t) e^{i(\mathbf{k}_e \cdot \mathbf{r} - \omega t)}$ and $\mathbf{E}_p(\mathbf{r},t) = (1/\sqrt{2})(\hat{\mathbf{x}}+\hat{\mathbf{y}}) E_p(\mathbf{r},t-t_d) e^{i(\mathbf{k}_p \cdot \mathbf{r} - \omega t)}$, where $t_d$ is the pump–probe time delay, the intensity is, in the limit of a weak probe,

$$I(\mathbf{r},t) = \frac{n_0 c}{8\pi}[|E_e(\mathbf{r},t)|^2 \\ + \frac{1}{\sqrt{2}}(E_e^*(\mathbf{r},t) E_p(\mathbf{r},t-t_d) e^{i(\mathbf{k}_p - \mathbf{k}_e)\cdot\mathbf{r}} + c.c.)] \quad (1)$$

The probe propagation direction is taken to be $\hat{\mathbf{z}}$. For $N_e \ll N_c$, the plasma contribution to the refractive index is approximated as $\Delta n_{pl}(\mathbf{r},t) = -N_e(\mathbf{r},t)/(2N_c)$, where $N_e(\mathbf{r},t)$ is the free electron density and $N_c$ is the critical density. In the intensity range studied in [1], multiphoton absorption determines the rate at which free electrons are generated $\dot{N}_e = \sigma_m I^m(t) N_0$ where $\sigma_m$ is the cross section for $m$–photon absorption and $N_0$ is the ambient gas density. The precise value of the exponent $m$ is not important for the conclusions drawn later – in the calculations we assume $m = 8$, which has been used previously for air for pulses centered at 800 nm in this intensity regime [16]. The ionization rate is, in the limit of a weak probe beam, $\dot{N}_e(\mathbf{r},t) = \dot{N}_e^s(\mathbf{r},t) + \dot{N}_e^g(\mathbf{r},t)$ where $\dot{N}_e^s(\mathbf{r},t) = \sigma_m I_e^m(\mathbf{r},t) N_0$ and

$$\dot{N}_e^g(\mathbf{r},t) = \dot{N}_e^s(\mathbf{r},t) \frac{m}{\sqrt{2}} \frac{E_e^*(\mathbf{r},t) E_p(\mathbf{r},t) e^{i(\mathbf{k}_p - \mathbf{k}_e)\cdot\mathbf{r}} + c.c.}{|E_e(\mathbf{r},t)|^2} \quad (2)$$

where the superscripts $s$ and $g$ refer to "smooth" and "grating". The interference between pump and probe beams causes a spatial modulation (grating) in the plasma density. A plasma grating has been shown to enable energy exchange between femtosecond laser filaments [17] – here we find it contributes a phase shift to the polarization component of the probe beam parallel to the pump polarization.

The smooth and grating parts of the refractive index change due to the plasma are

$$\Delta n_{pl}(\mathbf{r},t) = \Delta n_{pl}^s(\mathbf{r},t) + [\Delta n_{pl}^g(\mathbf{r},t) e^{i(\mathbf{k}_p - \mathbf{k}_e)\cdot\mathbf{r}} + c.c.], \quad (3)$$

$$\Delta n_{pl}^s(\mathbf{r},t) = -\frac{N_0}{2N_c}\int_{-\infty}^{t}\sigma_m I_e^m(\mathbf{r},t')dt', \quad (4)$$

and

$$\Delta n_{pl}^g(\mathbf{r},t) = -\frac{n_0 c N_0}{2\sqrt{2}\pi N_c} \quad (5)$$

$$\times \int_{-\infty}^{t}\sigma_m I_e^{m-1}(\mathbf{r},t')E_e^*(\mathbf{r},t')E_p(\mathbf{r},t'-t_d)dt'$$

Using an instantaneous Kerr nonlinearity $n_2$ for Ar, the change in refractive index due to the Kerr effect is $\Delta n_K(\mathbf{r},t) = n_2 I(\mathbf{r},t)$. In analogy with the discussion above, the refractive index change due to the Kerr effect is $\Delta n_K(\mathbf{r},t) = \Delta n_K^s(\mathbf{r},t) + [\Delta n_K^g(\mathbf{r},t)e^{i(\mathbf{k}_p-\mathbf{k}_e)\cdot\mathbf{r}} + c.c.]$, where $\Delta n_K^s(\mathbf{r},t) = n_2 I_e(\mathbf{r},t)$ and
$\Delta n_K^g(\mathbf{r},t) = n_2 n_0 c/(16\pi\sqrt{2})E_e^*(\mathbf{r},t)E_p(\mathbf{r},t-t_d)$.

We solve the wave equation

$$\nabla^2 \mathbf{E} - \frac{1}{c^2}\frac{\partial^2 \mathbf{D}}{\partial t^2} = 0 \quad (6)$$

When the pulse envelope changes slowly compared to an optical cycle, as is the case here, one can write $\mathbf{D} = n^2\mathbf{E}$, and we use $n^2 \approx n_0^2 + 2n_0(\Delta n_{pl}+\Delta n_K)$ for small index shifts. The refractive index is a function of time due to the plasma and Kerr contributions to $n$. We define the following to enable full normalization of the propagation equation: $I_e = I_{e0}f(\mathbf{r},t)$, where $f$ is an intensity envelope function, $E_e = E_{e0}f^{1/2}$, $E_p = E_{p0}f^{1/2}$, $\Delta n_K = \Delta n_{K0}f$, $\Delta n_{K0} = n_2 I_{e0}$, $z' = k_p \Delta n_{K0} z$, $\tau' = \omega t - z'$, $\Delta n_{pl,0} = -\sigma_m N_0 I_{e0}^m t_w/(2N_c)$, $u = t/t_w$, $u_d = t_d/t_w$, $\xi = \Delta n_{pl,0}/\Delta n_{K0}$, and $\delta = E_{p0}/E_{e0}$. Here $k_p = n_0\omega/c$ and $t_w$ is a characteristic pump/probe pulse duration. As seen later, the probe phase shift per unit length of propagation is entirely determined by $\xi$, the characteristic ratio of plasma to Kerr response.

Employing the above definitions plus the slowly varying envelope and paraxial approximations, we get for the $j = x, y$ field components propagating in the direction of the probe $\mathbf{k}_p$

$$\frac{\partial E^j}{\partial z'} = E_p(i - 2\frac{\partial}{\partial\tau'})s_j - 2s_j\frac{\partial E_p}{\partial\tau'} \quad (7)$$
$$+\sqrt{2}E_e(i-2\frac{\partial}{\partial\tau'})g_j - 2\sqrt{2}g_j\frac{\partial E_e}{\partial\tau'},$$

with $s_y = (\Delta n_K^s/3 + \Delta n_{pl}^s)n_0\Delta n_{K0}^{-1} = f/3 + \xi v(u)$,

$s_x = (\Delta n_K^s + \Delta n_{pl}^s)n_0\Delta n_{K0}^{-1} = s_y + 2f/3$

$g_y = n_0\Delta n_{K0}^{-1}\Delta n_K^g/3 = \delta f^{1/2}(u)f^{1/2}(u-u_d)/3$,

$g_x = (\Delta n_K^g + \Delta n_{pl}^g)n_0\Delta n_{K0}^{-1} = 3g_y + m\delta\xi h(u,u_d)$, where

$v(u) = \int_{-\infty}^{u}f^m(s)ds$ and $h(u,u_d) = \int_{-\infty}^{u}f^{m-1/2}(s)f^{1/2}(s-u_d)ds$.

The first two terms in Eq. (7), containing $s_j$, give the direct effect on the probe beam of the "smooth" part of the refractive index. The plasma contributes equally to both polarizations because its response is isotropic. The third and fourth terms, containing $g_j$, arise from coherent scattering (diffraction) of the pump pulse into the $\mathbf{k}_p$ direction by the Kerr and plasma gratings, which affects both the phase and amplitude of the probe field. The effect of the Kerr grating is implicitly taken into account by the cross phase modulation Kerr coefficient, and the same would apply for a higher-order Kerr effect [2]. But the phase shift imparted by the plasma grating has, to our knowledge, not been considered, at least in this context. The appearance of a grating term for the Kerr nonlinearity when the pump and probe are perpendicular is due to off-diagonal elements in the $\chi^{(3)}$ tensor. A plasma grating, however, is generated only when the pump and probe pulses interfere; there must be some polarization overlap.

To solve Eq. (7), we neglect energy transfer due to two-beam coupling [18-20] and concentrate on the phase shift imparted to the probe per unit interaction length. We neglect the time derivative terms in Eq. (7): they produce no probe phase shift to first order and are in any case small for 90 fs pulses. We note that the pump-probe crossing angle enters the calculation only through $z'$. The result is $\partial E^j/\partial z' = i\beta_j(u)E_p$, where $\beta_y = s_y + f/3$ and $\beta_x = 2f + \xi[v(u) + mh(u,u_d)f^{1/2}(u)f^{-1/2}(u-u_d)]$.

The phase shift per unit length of propagation of an $\hat{\mathbf{x}}$–polarized probe ($\beta_x(t)$) and a $\hat{\mathbf{y}}$–polarized probe ($\beta_y(t)$) are plotted versus pump-probe delay $t_d$ in Fig. 1(a). We assume 90 fs Gaussian pulses ($f(u) = \exp(-u^2)$) and $m = 8$. Note that these results are independent of the pump–probe crossing angle. For these curves, we turned off the Kerr response to highlight the effective birefringence of the plasma grating, which contributes only during pump–probe temporal overlap and for parallel polarizations. The peak phase shift due to the grating is a factor of $\sim m/2$ larger than the "smooth" plasma signal at $t_d \gg t_w$. Consistent with our normalized expressions, the phase shift is independent of $\delta = E_{p0}/E_{e0}$ until the probe intensity is comparable with the pump intensity, whereupon our perturbation treatment fails in any case. The physical reason for this is that the pump scattering into the $\mathbf{k}_p$ direction is proportional to the grating modulation depth, which for a weak probe is proportional to $E_p$.

In the heterodyne experiment of [1], the signal $S(t_d)$ is the probe retardance (the difference in phase shifts of the perpendicular polarization components) weighted by the probe pulse envelope. We simulate this as $S(t_d) \propto \int_{-\infty}^{\infty}(\beta_x(t) - \beta_y(t))I_p(t,t_d)dt$. Results of this calculation are plotted in Fig. 1(b) for a range of values of the plasma to Kerr response ratio, $\xi = \Delta n_{pl,0}/\Delta n_{K0} = -3.8\times10^{-2}$, $-0.19$, $-0.35$, $-0.54$ and $-1.1$. For $|\xi| < 3.8\times10^{-2}$, the positive Kerr signal dominates. As $|\xi|$ increases beyond this point, the plasma grating begins to contribute and the Kerr effect appears to saturate. At $\xi = -0.54$, the plasma grating pushes the signal negative for $t_d \approx 0$. The Kerr component of the signal, because it is a convolution of the pump and probe pulse shapes, is wider in time than the plasma grating signal. This leads to positive wings in an intensity range that contains a strong negative peak as seen here and in the Loriot et al. results (see Fig. 7 in [2]). At $\xi = -1.1$, the plasma grating dominates.

To compare our simulations directly to the results of Loriot *et al.* [1], we evaluate $\xi$ for their conditions. From [1], we use $n_2 = 3 \times 10^{-19}$ cm$^2$/W, $I_{e0} = 1-22$ TW/cm$^2$ (the low end of their pump intensity range), $t_w = 54$ fs (corresponding to 90 fs full width at half maximum Gaussian), $N_c = 1.7 \times 10^{21}$ cm$^{-3}$ (at 800 nm), ambient gas density $N_0 = 2.7 \times 10^{18}$ cm$^{-3}$, $m=8$, and $\sigma_8 \sim 3.7 \times 10^{-96}$ cm$^{16}$/W$^8$/s [16]. Over 1–22 TW/cm$^2$, $\xi$ ranges from $-5.4 \times 10^{-10}$ to $-1.3$, overlapping the full range of our curves plotted in Fig. 1(b).

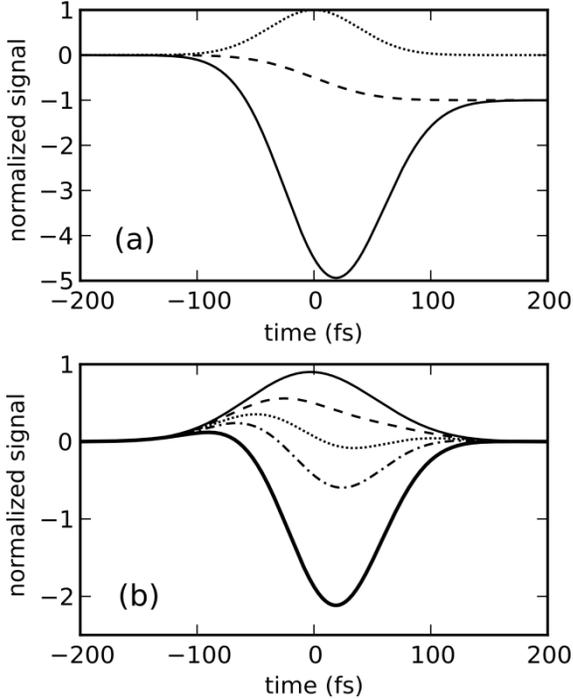

Fig. 1. Simulated normalized signal in a transient birefringence pump-probe experiment near the ionization threshold. (a) Effect of plasma grating: phase shift per unit interaction length for the component of the probe beam polarized parallel (solid) and perpendicular (dashed) to the pump polarization. The dotted curve shows the pulse envelope. (b) Transient birefringence at $\Delta n_{pl,0}/\Delta n_{K0} = -3.8 \times 10^{-2}$ (thin solid), $-0.19$ (dashed), $-0.35$ (dotted), $-0.54$ (dash-dot), and $-1.1$ (thick solid).

When the pump and probe pulses have very different carrier frequencies, the ionization rate oscillates in time as well as in space. Because the plasma density accumulates over the pulse, the spatial grating, and thus the coherent signal from the plasma, is suppressed. This is why the plasma grating effect is not observed when a nondegenerate probe is used [15,21]. These calculations show that the zero time delay signal in transient birefringence measurements cannot be assumed to arise from the Kerr effect alone when the pump intensity is near the ionization threshold. In particular, they cast new doubt on the high order Kerr effect-based interpretation of experiments [1-3], particularly when the body of evidence from other experiments is also considered [8-15,21]. Finally, we note that the grating effect associated with a delayed nonlinearity is present in other experimental techniques that are sensitive to the probe phase shift, such as cross defocusing [21, 22] and 4*f* coherent imaging [23].

J.K.W. acknowledges support from the Joint Quantum Institute. This research was supported by the National Science Foundation, the U.S. Department of Energy, the Office of Naval Research, and the Lockheed Martin Corporation.